\documentclass[12pt,authoryear]{elsarticle}
\usepackage{caption}
\usepackage{natbib}
\usepackage{graphicx,psfrag,epsfig}
\usepackage{amssymb, amsmath}
\usepackage{epstopdf}
\usepackage{url}
\usepackage{mathtools}
\usepackage{color}
\usepackage[mathscr]{euscript}

\usepackage{etoolbox} 
\makeatletter
\patchcmd{\ps@pprintTitle}{\footnotesize\itshape
       Preprint submitted to \ifx\@journal\@empty Elsevier
       \else\@journal\fi\hfill\today}{\relax}{}{}
\makeatother

\definecolor{Mygrey}{gray}{0.35}

\def\beginmat{ \left( \begin{array} }
\def\endmat{ \end{array} \right) }

\begin{document}

\begin{frontmatter}

\title{Covariance models on the surface of a sphere: when does it matter?}

\author{Jaehong Jeong}
\ead{jeong@stat.tamu.edu}
\author{Mikyoung Jun\corref{cor1}}
\ead{mjun@stat.tamu.edu}
\address{Department of Statistics, Texas A\&M University, College station, TX}

\cortext[cor1]{Corresponding author}

\begin{abstract}

There is a growing interest in developing covariance functions for processes on the surface of a sphere due to wide availability of data on the globe. Utilizing the one-to-one mapping between the Euclidean distance and the great circle distance,  isotropic and positive definite functions in a Euclidean space can be used as covariance functions on the surface of a sphere. This approach, however, may result in physically unrealistic distortion on the sphere especially for large distances. We consider several classes of parametric covariance functions on the surface of a sphere, defined with either the great circle distance or the Euclidean distance, and investigate their impact upon spatial prediction. We fit several isotropic covariance models to simulated data as well as real data from NCEP/NCAR reanalysis on the sphere. We demonstrate that covariance functions originally defined with the Euclidean distance may not be adequate for some global data.

\vspace{.3cm}
\end{abstract}


\end{frontmatter}


\vspace{.5cm}
\section{Introduction}

In geophysical and environmental sciences, data often come in a global scale. Covariance functions for global data sets need to be positive definite on the surface of a sphere and the distance computation is important in spatial modeling. For an integer $d\geq 1$, define $\mathcal{S}_{r}^{d}=\{\mathbf{x}\in\mathbb{R}^{d+1}:||\mathbf{x}||=r \}$ to be a ($d$-dimensional) sphere with radius $r$, where $||\mathbf{x}||$ is the Euclidean norm of $\mathbf{x}\in\mathbb{R}^{d+1}$. We also define the great circle distance on $\mathcal{S}_{r}^{d}$ by  $\theta(\mathbf{x},\mathbf{y})=r\times\arccos(\langle \mathbf{x},\mathbf{y}\rangle)$ where $\langle\cdot,\cdot\rangle$ denotes the usual inner product on $\mathbb{R}^{d+1}$. 

We consider the surface of the Earth as the spatial domain. Let $\mathcal{S}_{R}^{2}$  denote the surface of the Earth, where $R$ denotes the Earth's radius (we approximate the Earth as a perfect sphere). 
The chordal distance between the two points, $\mathbf{s}_{1}$=($L_{1}$,$l_{1}$) and $\mathbf{s}_{2}$=($L_{2}$,$l_{2}$), on $\mathcal{S}_{R}^2$ ($L$ and $l$ denote latitude and longitude, respectively) is given by
\begin{equation*}
{\rm ch}(\mathbf{s}_{1},\mathbf{s}_{2})=2R[{\rm sin}^{2}\{(L_{1}-L_{2})/2\}+{\rm cos}L_{1}{\rm cos}L_{2}{\rm sin}^{2}\{(l_{1}-l_{2})/2\} ]^{1/2}.
\end{equation*} 
 The great circle distance between the two locations then is given by $\theta=\theta(\mathbf{s}_{1},\mathbf{s}_{2})=2R\times\arcsin\{{\rm ch}(\mathbf{s}_{1},\mathbf{s}_{2})/(2R)\}$. The chordal distance is simply the Euclidean distance penetrating the spatial domain on the surface of the Earth, and producing a straight line approximation to the great circle distance \citep{banerjee2005geodetic}. 

\cite{yadrenko1983spectral} pointed out that any covariance function in $\mathbb{R}^{3}$ can be considered as a covariance function for processes on $\mathcal{S}_{r}^{2}$ using the chordal (i.e. Euclidean) distance. This construction can provide a rich class of covariance functions on $\mathcal{S}_{r}^{2}$ \citep{gneiting1999correlation}. As argued in \cite{gneiting2013strictly}, the great circle distance is a physically most natural distance metric for processes on a sphere. However, literature on covariance modeling using the great circle distance on the surface of a sphere is scarce due to its mathematical challenge. Some efforts have been made in examining the validity of several parametric covariance functions on the surface of a sphere \citep{huang2011validity,gneiting2013strictly} and in developing valid parametric covariance functions with the great circle distance from various constructional approaches \citep{du2012variogram,du2013isotropic,gneiting2013strictly,guinness2013covariance,jeong2015class}.

Although \cite{huang2011validity} and \cite{gneiting2013strictly} studied validity of covariance functions defined with either the great circle distance or the Euclidean distance in details, the impact upon parameter estimation and prediction has not been studied well. According to \cite{banerjee2005geodetic}, careless formulation of distances can lead to poor prediction with wrong estimation of the spatial range. Note that this study considered the Mat\'{e}rn covariance model using the great circle distance, which may not be positive definite on the surface of a sphere, unless the smoothness parameter, $\nu$, is $\nu\in(0,0.5]$ \citep{miller2001completely,gneiting2013strictly}.

In this paper, we consider several positive definite functions on $\mathbb{R}^{d+1}$ and $\mathcal{S}_{r}^{d}$ for $d=1,2$, and compare them in simulation studies and real data applications.The rest of the paper is organized as follows. In Section~\ref{sec:char}, we discuss some characteristics of covariance functions on a sphere. Then we present two simulation studies on $\mathcal{S}_{1}^{1}$ and $\mathcal{S}_{R}^{2}$ in Section~\ref{sec:sim}. Section~\ref{sec:app} illustrates real application results to geopotential height data set. Finally, Section~\ref{sec:dis} concludes the paper with discussion.

\vskip14pt
\section{Characteristics of covariance functions on a sphere}
\label{sec:char}

We first review some known results on covariance functions in the Euclidean space as well as those on the surface of a sphere. A function $f:\mathbb{R}^{d}\times\mathbb{R}^{d}\rightarrow\mathbb{R}$ is called positive definite if
\begin{equation}
\label{eq:pdf}
\hbox{$\sum_{i,j=1}^{n}$}c_{i}c_{j}f(\mathbf{x}_{i},\mathbf{x}_{j})\geq 0
\end{equation}
for all finite $n\in \mathbb{N}$, all distinct points $\mathbf{x}_{1},\dots,\mathbf{x}_{n}\in\mathbb{R}^{d}$, and all real $c_{1},\dots,c_{n}$. A function is strictly positive definite when the inequality in \eqref{eq:pdf} is strict (unless $c_1=c_2=\cdots=c_n=0$). For a real random field $Z$ in $\mathbb{R}^{d}$ with $E\{Z(\mathbf{x})\}^{2}<\infty$ for all $\mathbf{x}\in\mathbb{R}^{d}$, the covariance function $K(\mathbf{x},\mathbf{y})={\rm cov}\{Z(\mathbf{x}),Z(\mathbf{y})\}$ must satisfy the condition in (1). The random field $Z$ is called weakly stationary if its mean function is constant, it has finite second moments, and its covariance function can be written as ${\rm cov}\{Z(\mathbf{x}),Z(\mathbf{y})\}=K(\mathbf{x}-\mathbf{y})$ for all $\mathbf{x},\mathbf{y}\in\mathbb{R}^{d}$, and a positive definite function $K$, i.e. the covariance function of $Z$, depends on $\mathbf{x}$ and $\mathbf{y}$ only through $\mathbf{x}-\mathbf{y}$. Furthermore, if its covariance function satisfies ${\rm cov}\{Z(\mathbf{x}),Z(\mathbf{y})\}=\varphi(||\mathbf{x}-\mathbf{y}||)$ for a positive definite function $\varphi$, then the random field $Z$ is weakly isotropic. An isotropic property for processes in $\mathbb{R}^{d}$ can be thought as an invariance property under translation and rotation  \citep{stein1999interpolation}. 

The covariance function of a random field and the smoothness of its realization are related to mean square properties of the random field.  The random field $Z$ is called mean square continuous at $\mathbf{x}$ if
\begin{equation*}
E\{Z(\mathbf{y})-Z(\mathbf{x})\}^{2}\rightarrow 0\ \ {\rm as}\ \mathbf{y}\rightarrow \mathbf{x}.
\end{equation*}
For weakly stationary random field, mean square continuity is equivalent to the fact that the covariance function is continuous at the origin, but it does not imply continuity of its realization \citep{stein1999interpolation}. Moreover, a random field $Z$ on $\mathbb{R}$ with finite second moments is mean square differentiable at $t$ if there exists $Z'(t)=\hbox{$\lim_{n\rightarrow\infty}$}\{Z(t+h_{n})-Z(t)\}/h_{n}$  in $L^{2}$ for sequences $h_{n}\rightarrow 0$. The smoothness of a random field can be determined through the number of mean square derivatives. \cite{gneiting2013strictly} defined the class of $\Phi_{d}$ with the correlation functions of mean square continuous, stationary and isotropic random fields in $\mathbb{R}^{d}$. Every positive definite function $\varphi:[0,\infty)\rightarrow\mathbb{R}$ with $\varphi(0)=1$ is the correlation of an isotropic process and the members of $\Phi_{2}$ and $\Phi_{3}$ are the cornerstones for covariance models for spatial data in a planar domain \citep{gneiting2013strictly}.
An isotropic property on a sphere means the covariance function depends on distance only. That is, a random field $Z$ on $\mathcal{S}_{r}^{d}$ is called isotropic if its covariance function satisfies ${\rm cov}\{Z(\mathbf{x}),Z(\mathbf{y})\}=\psi(\theta(\mathbf{x},\mathbf{y}))$ for all $\mathbf{x},\mathbf{y}\in\mathcal{S}_{r}^{d}$.
We then similarly define $\Psi_{d}$,  the class of continuous, isotropic covariance functions $\psi:[0,\pi \times r] \rightarrow \mathbb{R}$ on $\mathcal{S}_{r}^{d}$. 

Since a sphere can be viewed as a subset of the Euclidean space, valid covariance functions on $\mathbb{R}^{d+1}\times\mathbb{R}^{d+1}$ can be restricted to $\mathcal{S}_{r}^{d}\times\mathcal{S}_{r}^{d}$ when the Euclidean distance is used (equivalently, the chordal distance on $\mathcal{S}_{r}^{d}$). \cite{yadrenko1983spectral} and \cite{yaglom1987correlation} pointed out that if $\varphi$ is a member of the class $\Phi_{d+1}$, then the function $\varphi[2r\sin\{\theta/(2r)\}]$, with the Euclidean distance expressed in terms of great circle distance as $2r\sin\{\theta(\mathbf{x},\mathbf{y})/(2r)\}$, belongs to the class $\Psi_{d}$. Since there are various positive definite functions, including the Mat\'{e}rn class and the generalized Cauchy families \citep{stein1999interpolation,gneiting2013strictly} that are  isotropic covariance functions for processes in $\mathbb{R}^{3}$, this mapping from $\varphi\in\Phi_{3}$ to $\psi\in\Psi_{2}$ provides a useful way to construct a rich parametric class of isotropic covariance functions on $\mathcal{S}_{r}^{2}$. This mapping preserves the interpretation of parameters such as scale, range, smoothness, and fractal index \citep{gneiting2013strictly}. 

It has been reported in the literature (e.g. \cite{guinness2013covariance}, \cite{jeong2015class}) that when Mat\'{e}rn class with the Euclidean distance and that with the great circle distance are compared in terms of model fit and prediction, often Mat\'{e}rn model with the Euclidean distance performs better. This may be due to the restriction on the smoothness parameter for the Mat\'{e}rn class with the great circle distance. \cite{jeong2015class} proposed a method to overcome such limitation on the smoothness parameter for the Mat\'{e}rn class with the great circle distance, but they found that the Mat\'{e}rn class with the Euclidean distance is equivalent or often better compared to the covariance models specifically developed for processes on the sphere. Our goal in this paper is to study cases that are not previously considered in the literature, and to explore cases where there are significant differences (improvements) of covariance models defined on the sphere as opposed to the covariance models projected from the Euclidean space.

We focus on the fact that there are some fundamental differences between covariance models originally defined on the surface of a sphere and those in the Euclidean space. For instance, there exists a lower bound on isotropic correlation function in the Euclidean space. A function $\varphi$ is an isotropic correlation function in $\mathbb{R}^{d}$  if and only if it has the form, $\varphi(t)=\int_{0}^{\infty}\Lambda_{d}(tu)dG(u)$, where $\int_{0}^{\infty}dG(u)=1$ and $G$ is nondecreasing. Note that $\Lambda_{d}(r)=2^{(d-2)/2}\Gamma(d/2)r^{-(d-2)/2} {J}_{(d-2)/2}(r)$ where ${J}$ is a Bessel function. Then, for all $t$,
\begin{equation}
\label{eq:bound}
\varphi(t)\geq \hbox{$\inf_{s\geq 0}$} \Lambda_{d}(s)
\end{equation} \citep{stein1999interpolation}. This implies that valid correlation functions on $\mathcal{S}_r^2$ constructed through the mapping described above from $\varphi\in\Phi_{3}$ cannot have values less than $\hbox{$\inf_{s\geq 0}$} s^{-1}\sin s=-0.218$. In particular, the Mat\'{e}rn class yields nonnegative correlations only. Although the importance of the Mat\'{e}rn family is highlighted by \cite{stein1999interpolation} because of its flexibility with regard to the local behavior of the processes, it might not be appropriate in applications where there is significantly negative spatial correlation. In fact, many of the isotropic covariance functions in $\mathbb{R}^d$ used in the literature take non-negative values only.

We also focus on the fact that on the sphere, correlation between two points large distance apart may not necessarily be small (compared to correlation between nearby two points). In fact, if there is a wave traveling around the sphere, two points nearly maximum possible distance apart may be perfectly positively (or negatively) correlated, which cannot happen in the Euclidean space.

We now list several parametric classes of covariance functions defined on the surface of a sphere, or defined originally in the Euclidean space (then can be used through the projection). Some of the models are used in the simulation and data examples. We only consider isotropic covariance models on a sphere. 
Functions in the class $\Psi_{d}$ are characterized in terms of an infinite sum of Gegenbauer polynomials with nonnegative coefficients and cosine of the great circle distance \citep{schoenberg1942positive,gneiting2013strictly}. For $d\geq 1$, the class $\Psi_{d}$ consists of the functions of the form
\begin{equation*}
\psi(\theta)=\hbox{$\sum_{n=0}^{\infty}$} b_{n,d}C_{n}^{(d-1)/2}(\cos(\theta/r))/C_{n}^{(d-1)/2},\ \ \theta\in[0,\pi\times r],
\end{equation*} 
with nonnegative coefficients $b_{n,d}$ such that $\hbox{$\sum_{n=0}^{\infty}b_{n,d}$}=1$ and the Gegenbauer polynomial of degree n, $C_{n}^{(d-1)/2}$ \citep{schoenberg1942positive,chen2003necessary}. Moreover, the class $\Psi_{\infty}$ consists of the functions with the following form
\begin{equation}
\label{eq:psi_form}
\psi(\theta)=\hbox{$\sum_{n=0}^{\infty}$}b_{n}(\cos(\theta/r))^{n}, \ \ \theta\in[0,\pi\times r],
\end{equation}
with nonnegative coefficients $b_{n}$ such that $\hbox{$\sum_{n=0}^{\infty}b_{n}$}=1$.
The infinite sum is strictly positive definite on $\mathcal{S}_{r}^{d}$ when the coefficients $b_{n,d}$ and $b_{n}$ are strictly positive for infinitely many odd and infinitely many even integers $n$, and only a few closed forms, such as the multiquadratic family, for such infinite sums are known in general. 
The multiquadratic covariance function \citep{gneiting2013strictly} is defined by
\begin{equation*}
\psi(\theta)=\sigma^{2}(1-\tau)^{2c}/\{1+\tau^{2}-2\tau\cos(\theta/r) \}^{c},\ \ \theta\in[0,\pi\times r]
\end{equation*}
from a standard Taylor series of \eqref{eq:psi_form}, when $\sigma^{2}>0$, $c>0$, and $\tau\in(0,1)$. 
The Mat\'{e}rn class, given as 
\begin{equation}
\label{eq:matern}
\varphi(t)=\sigma^{2}{2^{\nu-1}}{\Gamma(\nu)}^{-1}(t/\alpha)^{\nu}\mathcal{K}_{\nu}(t/\alpha),\ \ t\geq 0,
\end{equation}
where the parameters, $\sigma^{2},\alpha,\nu>0$, are  marginal variance, spatial range, and smoothness parameters, respectively, is positive definite in $\mathbb{R}^{d}$ for any $d\in\mathbb{N}$ with the Euclidean distance \citep{stein1999interpolation}. Here, $\mathcal{K}_{\nu}$ is the modified Bessel function of the second kind of order $\nu$. 

\cite{gneiting2013strictly} showed that completely monotone functions (that have derivatives $\varphi^{(k)}$ of all orders with $(-1)^{k}\varphi^{(k)}(t)\geq 0$ for all nonnegative integers $k$ and all positive $t$) including the power exponential,  Mat\'{e}rn, and generalized Cauchy families are positive definite, through the restriction of a function $\varphi:[0,\infty)\rightarrow\mathbb{R}$ to the interval $[0,\pi\times r]$: $\psi=\varphi_{[0,\pi\times r]}$ under applicable conditions, on $\mathcal{S}_{r}^{d}$ of any dimension. One necessary condition of the membership in the class $\Psi_{d}$ is that either the fractal index or the smoothness parameter requires to satisfy $\beta\in(0,1]$ or $\nu\in(0,0.5]$, respectively. The powered exponential family defined by
\begin{equation*}
\psi(\theta)=\sigma^{2}{\rm exp}[-\{\theta/(cr)\}^{\beta}],\ \ \theta\in[0,\pi\times r]
\end{equation*}
where $\sigma^{2}>0$, $c>0$ is valid on any dimensional $\mathcal{S}_{r}^{d}$ if $\beta\in(0,1]$, and the Mat\'{e}rn family requires $\nu\in(0,0.5]$ similarly.
\cite{jeong2015class} compare Mat\'{e}rn class with great circle distance to that the Euclidean distance for spatial data on the surface of a sphere. 

Compactly supported members of the class $\Phi_{3}$ may be valid on $\mathcal{S}_{r}^{2}$  through $\psi=\varphi_{[0,\pi\times r]}$ \citep{gneiting2013strictly}. The spherical and the Wendland's functions on $\Phi_{3}$, remain valid with direct substitution of the Euclidean distance by the great circle distance on $\Psi_{d}$, $d=1,2,3$. In particular, the $C^{4}$-Wendland covariance function \citep{wendland1995piecewise}, defined as
\begin{equation}
\label{eq:wend}
\psi(\theta)=\sigma^{2}\{ 1+({\theta}\tau)/({cr})+{\theta^{2}}({\tau^{2}-1})/({3}{c^{2}r^{2}}) \}\{1-{\theta}/({cr})\}_{+}^{\tau},\ \ \theta\in[0,\pi\times r],
\end{equation}
where $c\in(0,\pi]$ is a support parameter and $\tau\geq 6$ is a shape parameter, has 4 derivatives at the origin and thus may be suitable for smooth data on $\mathcal{S}_r^d$.

The following covariance function (defined in the Euclidean space)  models a hole effect,
\begin{equation}
\label{eq:hole}
\varphi(t)=\sigma^{2}({\alpha}/{t})\sin(t/{\alpha}),\ \ t>0,
\end{equation}
with $\varphi(0)=\sigma^{2}$, and it is called the wave covariance function. This function may be deal with the situation where correlation between two points far apart may have bigger (in magnitude) than correlation between two points closer, or oscillating pattern in the correlation function. This belongs to the class $\Phi_{d}$ for $d=1,2,3$, but it is not valid on $\mathcal{S}_{r}^{2}$ \citep{huang2011validity}. Thus, this function can only be applied to the data on the sphere through the projection from the Euclidean space.

\cite{schoenberg1942positive} noted that the class $\Psi_{d}$ enjoys the useful closure properties. The class $\Psi_{d}$ is convex, closed under products, and closed under limits which are continuous. For examples, if $\psi_{1}(\theta),\psi_{2}(\theta)\in\Psi_{d}$, then $\lambda\psi_{1}(\theta)+(1-\lambda)\psi_{2}(\theta)\in\Psi_{d}$ for every $\lambda\in[0,1]$  and $\psi_{1}(\theta)\times\psi_{2}(\theta)\in\Psi_{d}$. Moreover, if $\psi_{n}(\theta)\in\Psi_{d}$, $\psi_{n}(\theta)\rightarrow\psi(\theta)$ as $n\rightarrow\infty$, and $\psi(\theta)$ is continuous, then also $\psi(\theta)\in\Psi_{d}$. These closure properties of the class $\Psi_{d}$ offer additional flexibility to model negative correlations. If we use convex sums or products of parametric families of correlation functions on $\mathcal{S}_{r}^{d}$ in terms of the great circle distance with a Legendre function of the form $\psi(\theta)=C_{n}^{(d-1)/2}\cos(\theta/r)/C_{n}^{(d-1)/2}$, including the case $\cos(\theta/r)$ for $d=1,2,3$, we can easily model negative correlations \citep{gneiting2013strictly}.

The sine-power covariance function is defined by
\begin{equation}
\psi(\theta)=\sigma^{2} \{1-(\sin\hbox{$\frac{\theta}{2r}$})^{\beta}\},\ \ \theta\in[0,\pi\times r],
\end{equation}
where the parameter $\beta\in(0,2]$ corresponds to the fractal index and controls the smoothness of the process. This covariance function operates directly on a sphere (belongs to the class $\Psi_{d}$ for all dimensions). 
On the other hand, the cosine function, $\varphi(t)=\cos(nt)$, where $t\geq 0$, belongs to the class $\Phi_{1}$ only for any value of $n>0$. On $\mathcal{S}_{r}^{d}$, $\psi(\theta)=\cos(n\theta/r)$ for $\theta\in[0,\pi\times r]$, is non-strictly positive definite for all dimensions when $n=1$. For non-integer values of $n>0$, $\psi(\theta)$ is not valid on $\mathcal{S}_{r}^{1}$ and for integer values of $n\geq 2$, it is non-strictly positive definite only on $\mathcal{S}_{r}^{1}$, but not on $\mathcal{S}_{r}^{2}$ \citep{gneiting2013strictly}.

When there is no significant negative correlations, and when the spatial range is small, we do not expect that using the great circle distance or the Euclidean distance may make significant difference. Moreover, when the prediction location is surrounded by enough number of estimation locations, we do not expect significant difference between models defined on the sphere and models defined in the Euclidean space (and used through projection).
Therefore, for the rest of the paper, we mainly consider the situations where either spatial range is large, or when there are not many number of estimation locations close to the prediction locations. Such situation may arise in real application, for example, when we are interested in predicting over the ocean with most observations on the land, or when we are interested in predicting near the poles with not many observations near the poles. We also consider cases where there is significantly negative correlations at large distances. 

\vskip14pt
\section{Simulation studies}
\label{sec:sim}


We present two simulation studies on $\mathcal{S}_{1}^{1}$ and $\mathcal{S}_{R}^{2}$. In the first example (Section~\ref{sec:exam_s1}), the truth is generated from an exponential covariance model with the great circle distance. In the second example (Section~\ref{sec:exam_s2}), the truth is generated from oscillating Mat\'{e}rn covariance model implemented in {\tt R-INLA} package \citep{martins2013bayesian}.

\subsection{Example on $\mathcal{S}_{1}^1$}
\label{sec:exam_s1}

\begin{figure}
\begin{center}
\includegraphics[height=3.25in,width=3.25in]{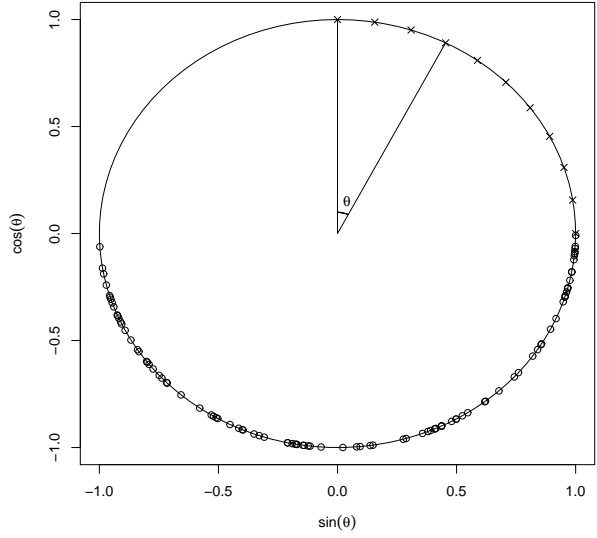}
\caption{\label{fig:circle_data}(Simulation example on $\mathcal{S}_{1}^1$) A realization of sampling locations ($\circ$) and prediction locations ($\times$). }
\end{center}
\end{figure}

We simulated mean zero Gaussian random fields on $\mathcal{S}_{1}^{1}$ using exponential covariance models with various range parameters using the great circle distance. The exponential covariance function on $\mathcal{S}_{r}^{d}$ is defined by
\begin{equation*}
\psi(\theta)=\sigma^{2}{\rm exp}(-\theta/\alpha),\ \ \theta\in[0,\pi\times r],
\end{equation*}
where $\sigma^{2}>0$ is a variance and $\alpha>0$ is a spatial range parameter. We then compared fitted results using the true covariance model as well as exponential covariance model with the Euclidean distance, i.e. $\varphi(t)=\sigma^{2}{\rm exp}(-t/\alpha)$, $t>0$. Exponential covariance functions belong to both $\Phi_{2}$ and $\Psi_{1}$, so they are valid on the surface of a sphere regardless of distance. We set the marginal variance $\sigma^{2}=1$ and considered the various spatial ranges $\alpha=2\pi,1.5\pi,\pi,\pi/1.5,\pi/2,\pi/4$.
We randomly selected 100 locations (angles) for parameter estimation from $(\pi/2,3\pi/2)$ on a unit circle, and 10 fixed and equally spaced locations for prediction from $[0, \pi/2)$ as in Figure~\ref{fig:circle_data}. We compared the covariance models in terms of prediction using the root mean square error (RMSE) as well as the two popular scoring rules \citep{gneiting2007strictly}, the mean absolute error (MAE) and the continuous ranked probability score (CRPS). We repeated this experiment 100 times: sampling locations are different each time and prediction locations are fixed. 

\begin{figure}
\begin{center}
\includegraphics[height=3.25in,width=6.5in]{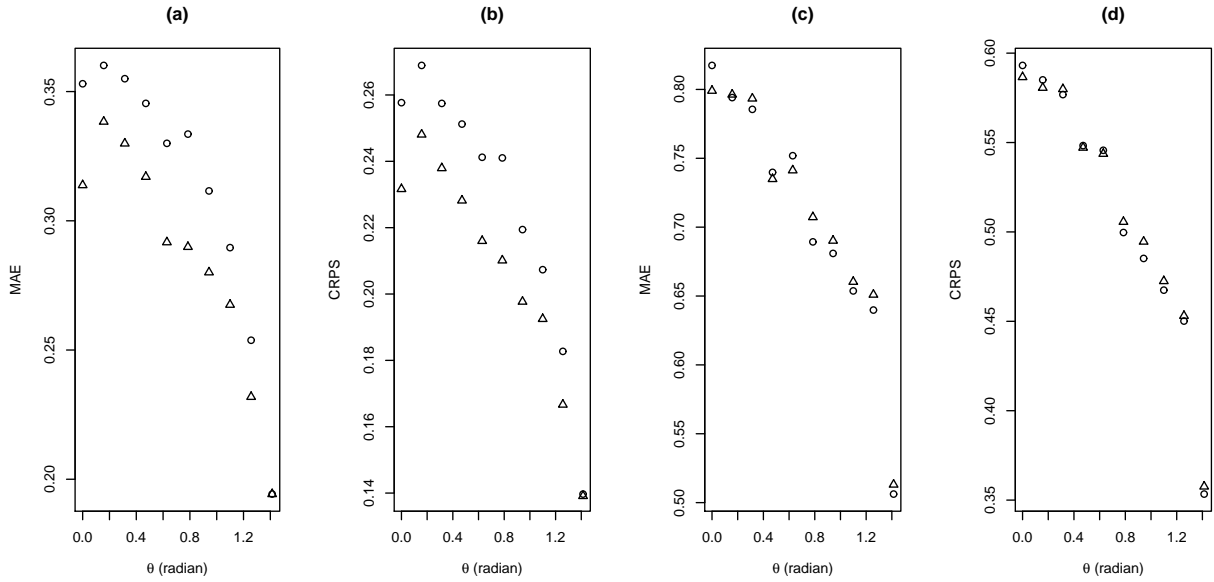}
\caption{\label{fig:circle_pm}(Simulation example on $\mathcal{S}_{1}^1$) MAE (a) and mean CRPS (b) averaged over 100 replications from the two models displayed against prediction locations ($\theta$) when the true spatial range $\alpha=2\pi$. (c) and (d) are the same as (a) and (b), except that $\alpha=\pi/4$. Triangles and circles represent the values of prediction errors for the exponential models using great circle and chordal distances, respectively.}
\end{center}
\end{figure}

Figures~\ref{fig:circle_pm}(a)-(d) present MAE and CRPS from the two models, the exponential models using great circle and chordal distances, displayed against 10 fixed prediction locations, for the true spatial range values are $2\pi$ and $\pi/4$. For the larger spatial range, $\alpha=2\pi$, we observe that the exponential model using the great circle distance performed significantly better than that using the chordal distance. Except for prediction locations that are relatively close to sampling locations, there are considerable differences between two models in prediction errors. On the other hand, for the smaller spatial range, $\alpha=\pi/4$, there is no significant difference between two models in both prediction errors, which agrees with findings in \cite{guinness2013covariance} and \cite{jeong2015class}. 

\subsection{Example on $\mathcal{S}_{R}^2$}
\label{sec:exam_s2}

\begin{figure}
\begin{center}
\includegraphics[height=3.25in,width=6.5in]{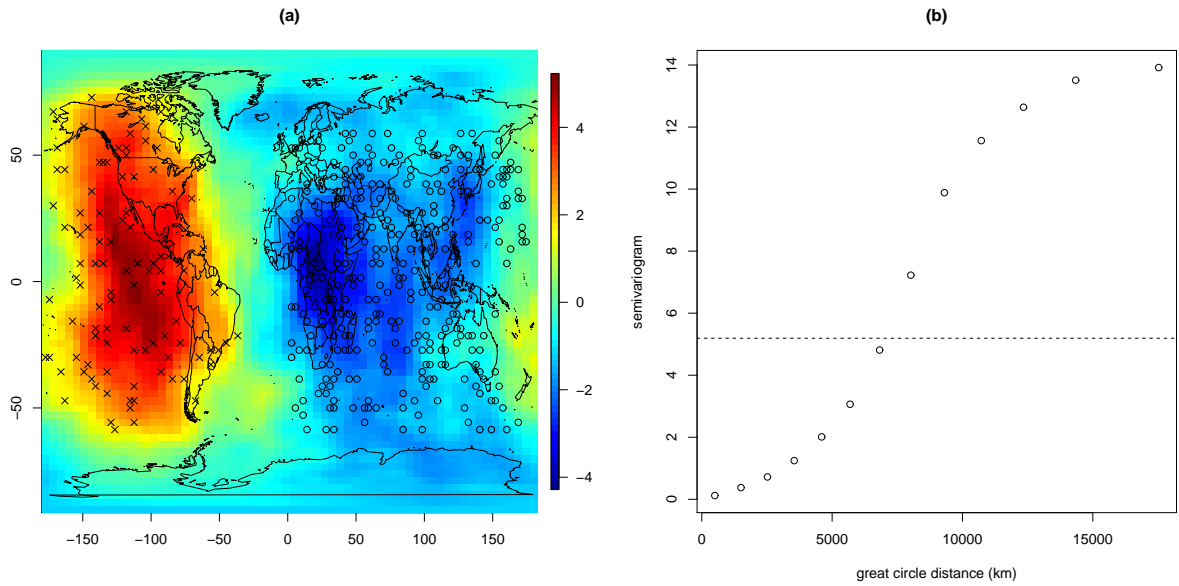}
\caption{\label{fig:osc_data}(Simulation example on $\mathcal{S}_{R}^2$) (a) A realization of residual fields. For (a), sampling locations ($\circ$) and prediction locations ($\times$). (b) Empirical semi variogram values for selected locations versus the great circle distance. For (b), dotted line represents sample variance.}
\end{center}
\end{figure}

We considered mean zero Gaussian random fields on the surface of the Earth (with radius $R=6,371$ (km)) with a Mat\'{e}rn covariance function from oscillating stochastic partial differential equations (SPDE) models with the {\tt spde1} class from the {\tt R-INLA} version 0.0-1413638221. We set $\sigma=1$, $\kappa=0.5$, $\tau=\kappa\times\sigma/\sqrt{4\pi}$, $\nu=2$ and $\theta_{osc}=0.3$. Here $\kappa$ is the spatial scale parameter, $\tau$ controls the variance with $\sigma$, $\nu$ controls the smoothness of the process, and $\theta_{osc}$ controls the strength of oscillation. We subtracted the constant mean (average over all locations) to have mean zero residual fields as in Figure~\ref{fig:osc_data}(a).
There are $128$ longitude points and $64$ latitude points, and the size of the data is $8,192$. We randomly selected $300$ locations where values are smaller than 0 for parameter estimation, and $100$ locations where values are larger than 1 for prediction. We repeated this procedure 100 times: all locations are different each time. It is clear from Figures~\ref{fig:osc_data}(a) and (b) that values at large distance lags are negatively correlated. 
We compared fitted results from a Mat\'{e}rn covariance model with the Euclidean distance (MC) to those from a Mat\'{e}rn covariance model with the great circle distance (MG). Moreover, to deal with negative correlations, we use convex sum of valid covariance functions with the great circle distance (C):
\begin{equation*}
\psi(\theta)=\sigma^{2}[\lambda\{1-(\sin{\hbox{$\frac{\theta}{2R}$}})^{\beta}\}+(1-\lambda)\cos(\theta/R)],\ \beta\in (0,2] \ {\rm and}\ \lambda\in (0,1),
\end{equation*}
We also considered the hole-effect model with the Euclidean distance (H) defined in \eqref{eq:hole}, a model defined with the Euclidean distance, for a comparison. 

\begin{figure}
\begin{center}
\includegraphics[height=3.25in,width=6.5in]{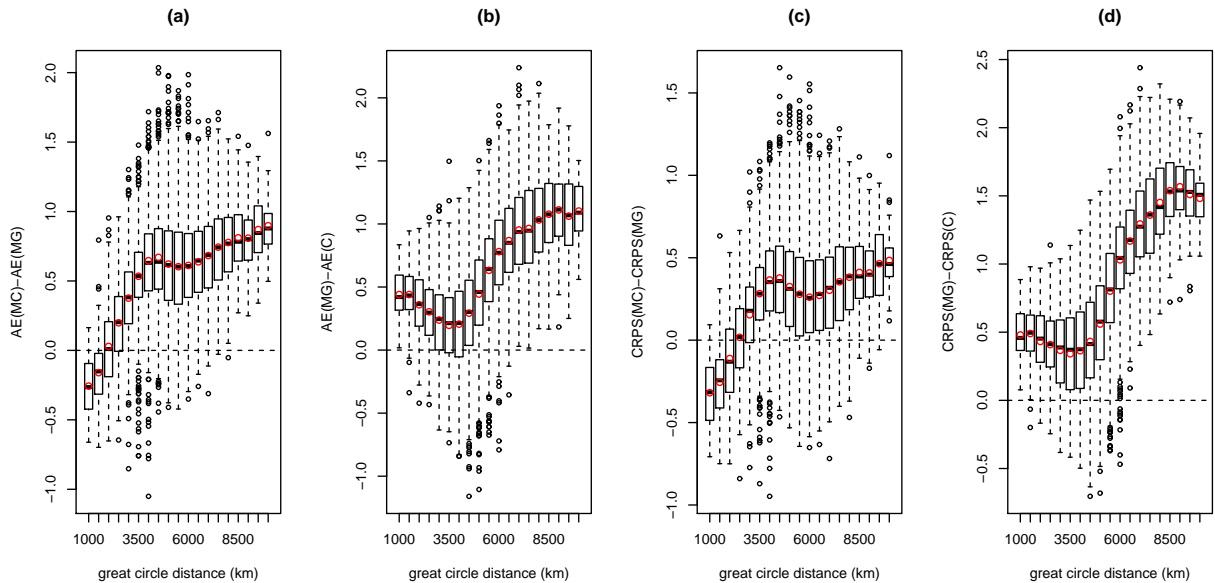}
\caption{\label{fig:osc_pm2}(Simulation example on $\mathcal{S}_{R}^2$) Boxplots of differences of AE (a) and CRPS (c) values from MC and MG, displayed against minimum great circle distance between a prediction location and its nearest sampling location. (b) and (d) The same as (a) and (c), except that selected models are MG and C. For (a)-(d), red circles represent average values in each bin.}
\end{center}
\end{figure}

Table~\ref{tab:s2} contains results of parameter estimation and prediction for the various models considered. From Table~\ref{tab:s2}, we observe that both MC and MG have large estimates of the spatial range parameter and MG has smaller prediction errors than MC. Although MC has better fit than MG in terms of maximum log-likelihood values, MC leads to poor prediction, possibly due to the large estimate of the spatial range. Note that the best model in terms of prediction errors is C. For C, the estimate of $\lambda$ is close to 1 and resulting estimated model is dominated by the sine-power model. Nevertheless, its correlation function allows much smaller correlation values for large distance lag, compared to the models for the Euclidean space. On the other hand, for H, although it allows negative correlations unlike Mat\'{e}rn covariance model, it resulted in poor model fit and spatial prediction. From Figures~\ref{fig:osc_pm2}(a)-(d), we observe significant differences of prediction errors between models using the great circle and Euclidean distances for prediction locations who are relatively far away from their nearest sampling locations. Overall, C outperforms MC and MG in prediction.

\vskip14pt
\section{Application}
\label{sec:app}

\subsection{Data and mean structure}

We consider geopotential height data on a global scale. The geopotential height approximates the actual height of a surface pressure at certain level above mean sea level. It is an adjustment to geometric height using the variation of gravity with elevation and latitude. The study of the geopotential height might be important in learning abnormal weather phenomena. According to the \cite{hafez2014recent}, the geopotential height at level 500 hPa plays a dominant role in controlling weather and climate conditions. Moreover, it became evident that the variability of global geopotential height is clearly impacted by global warming and climatic indices over the last several decades \citep{marshall2002trends,zhu2002global,hafez2012blocking,hafez2014recent}. 

Data sets are obtained at level 500 hPa from NCEP/NCAR reanalysis project ({\url {http://www.esrl.noaa.gov/psd/data/gridded/data.ncep.reanalysis.html}}). The NCEP/NCAR reanalysis 1 project uses a state of the art analysis/forecast system to perform data assimilation using past data from 1948 to the present (see \cite{kalnay1996ncep} for more detailed information). The output values are given on regular grids, and there are 144 longitude points and 73 latitude points (the size of the data is 10,512). We used Boreal summer geopotential height in the northern hemisphere (June, July, and August; JJA), and for each grid we computed a pointwise mean as the average over 2014. The unit is meter for the geopotential height and kilometer for distances. For variance stabilization, we took a square root transform of the data. 

\begin{figure}
\begin{center}
\includegraphics[height=3.25in,width=3.25in]{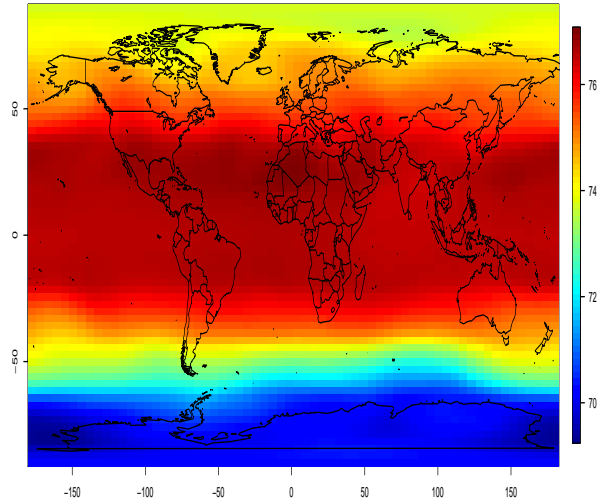}
\caption{\label{fig:geoT_data_vt}(Data example I) Square root of the geopotential height at level 500 hPa ($\sqrt{m}$). }
\end{center}
\end{figure}

We decompose the data into its mean structure (large scale variation) and the residual (for small scale spatial variation). Figure~\ref{fig:geoT_data_vt} suggests clear large scale spatial structure depending on latitude. Thus, we modeled the mean structure through simple harmonic regression:
\begin{align}
\label{eq:mlat2}
m(L) &= a_{0} + a_{1}\cos(L\times\pi/90^{\circ}) + a_{2}\sin(L\times\pi/90^{\circ}).
\end{align}
We considered two cases: one is with a constant (unknown) mean (that is, $a_{1}=a_{2}=0$ in \eqref{eq:mlat2}), and the other is given by \eqref{eq:mlat2}. In both cases, we first estimate the mean structure using regression and then work with the residual to fit the covariance structure.

\subsection{Example I: horizontal directional sampling design for prediction}
\label{sec:exam_hgt1}
\begin{figure}
\begin{center}
\includegraphics[height=3.25in,width=6.5in]{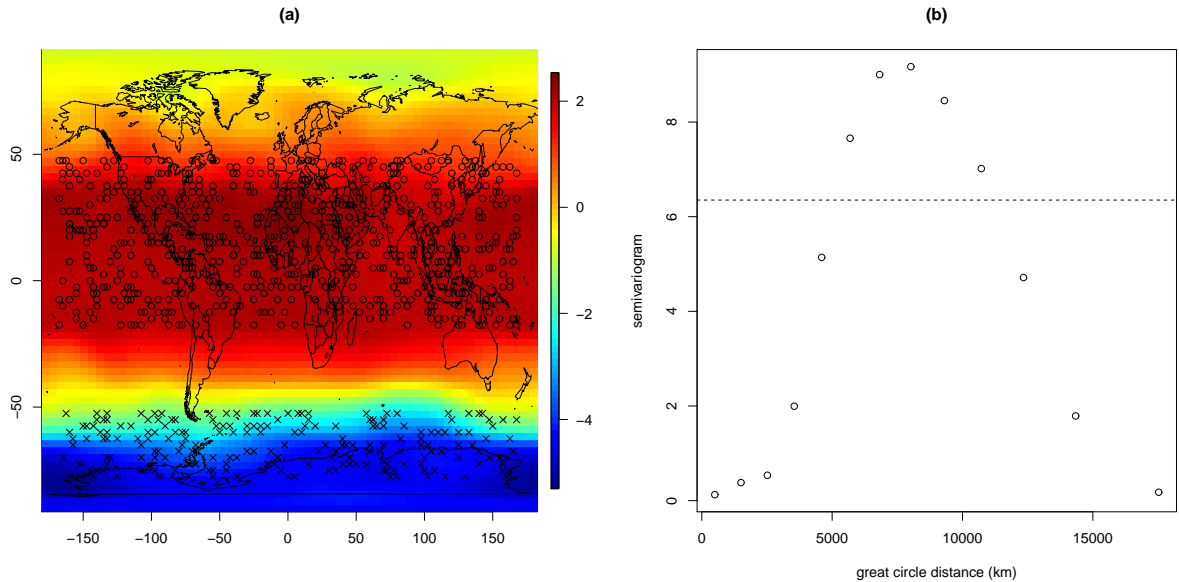}
\caption{\label{fig:geoT_data}(Data example I) (a) A realization of residual fields after subtracting the constant mean. For (a), sampling locations ($\circ$) and prediction locations ($\times$). (b) Empirical semi variogram values for selected locations versus the great circle distance. For (b), dotted line represents sample variance.}
\end{center}
\end{figure}

 Figure~\ref{fig:geoT_data}(a) shows the map  of residual after subtracting the constant mean and Figure~\ref{fig:geoT_data}(b) shows the empirical semi variogram of residuals. The semi variogram clearly shows negative covariances for large distances. To save computational burden, we randomly selected 600 locations near the red region for parameter estimation, and selected 200 locations over the blue region for prediction as in Figure~\ref{fig:geoT_data}(a). We repeated this process 100 times and for each time, all locations are randomly sampled and thus different. 
We compared the covariance models, MC, MG, and the convex sum model, C, with $C^4$-Wendland covariance function \eqref{eq:wend} and cosine function. As shown in Table~\ref{tab:geoT}, MG gives poor  model fit and prediction. Although MC has the largest maximum log-likelihood value, it leads to poor prediction compared to C. This is expected as the Mat\'{e}rn model is not able to produce negative correlations. Similarly to the simulation example of Section~\ref{sec:exam_s2}, C is the best model in terms of prediction.

\begin{figure}[htbp]
\begin{center}
\includegraphics[height=3.25in,width=6.5in]{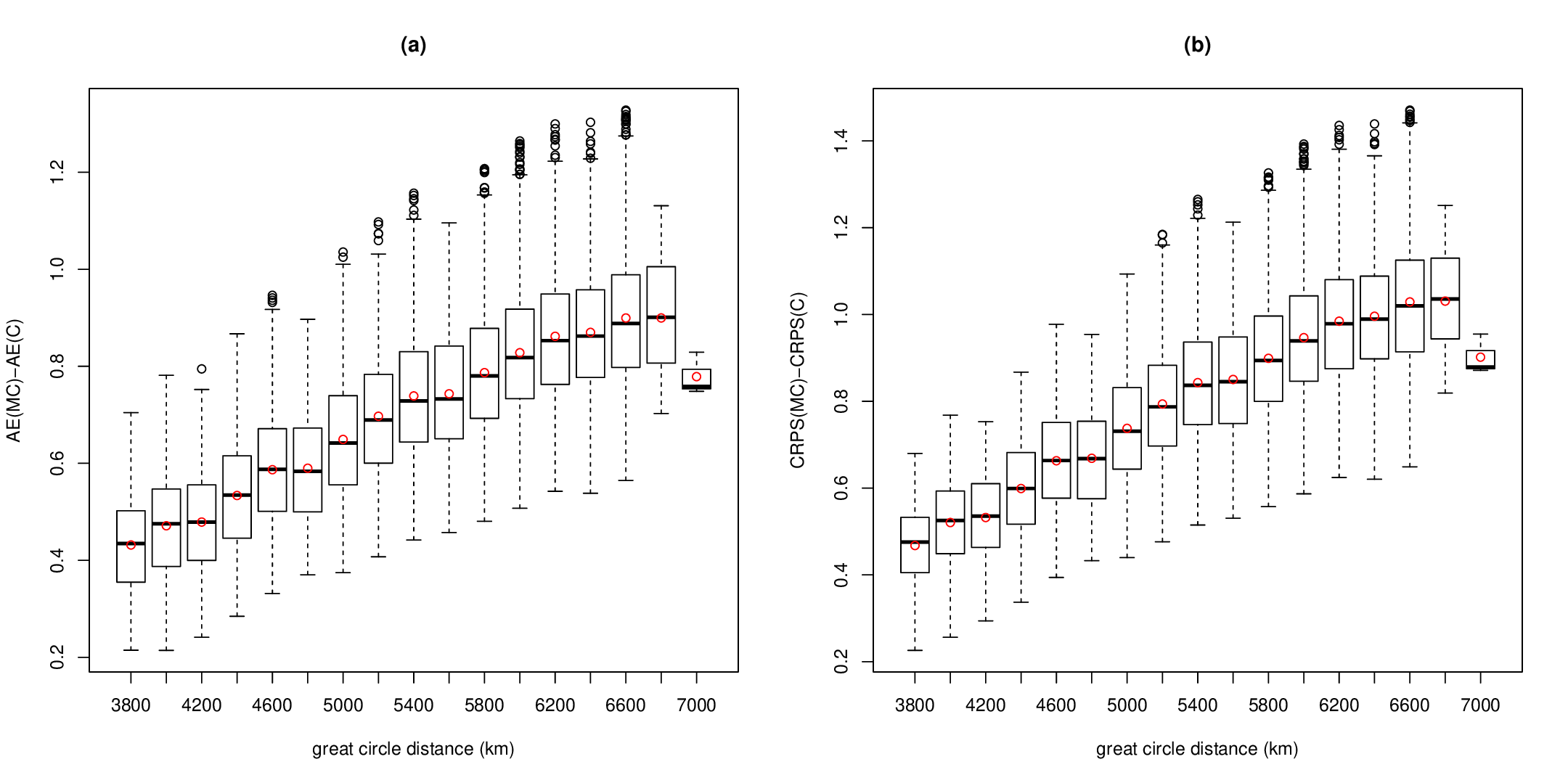}
\caption{\label{fig:geoT_pm3}(Data example I - the constant mean) Boxplots of differences of AE (a) and CRPS (b) values from MC and C, displayed against minimum great circle distance between a prediction location and its nearest sampling location. Red circles represent average values in each bin.}
\end{center}
\end{figure}

\begin{figure}
\begin{center}
\includegraphics[height=3.25in,width=6.5in]{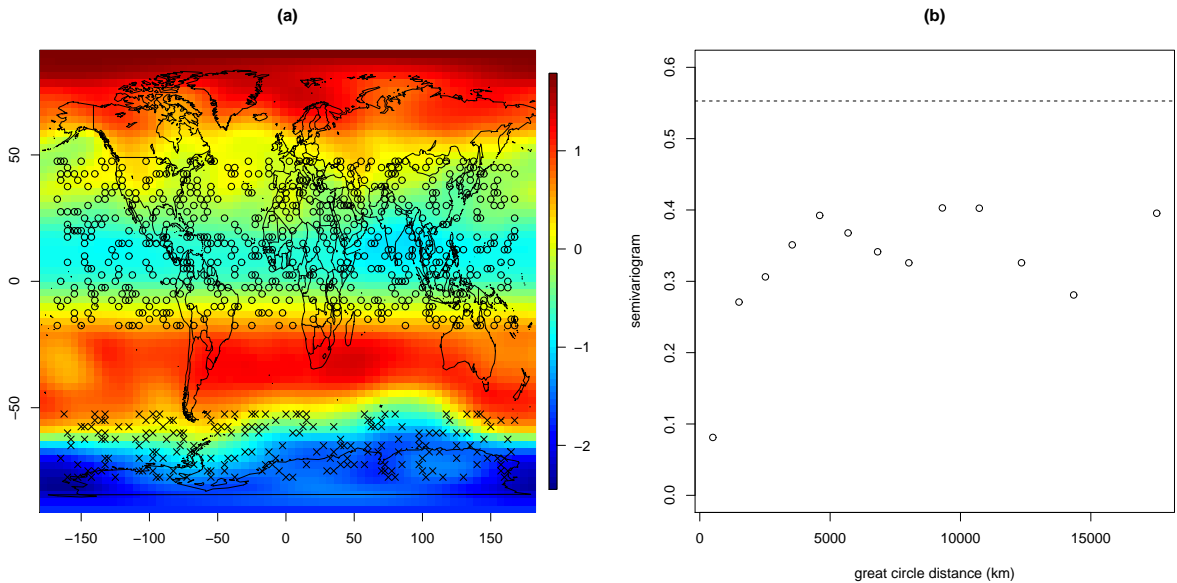}
\caption{\label{fig:geoT_LS_data}(Data example I) (a) A realization of residual fields after removing mean structure through simple harmonic regression depending on latitude. For (a), sampling locations ($\circ$) and prediction locations ($\times$). (b) Empirical semi variogram values for selected locations displayed against the great circle distance. For (b), dotted line represents sample variance. }
\end{center}
\end{figure}

Figures~\ref{fig:geoT_pm3}(a) and (b) show boxplots of differences of AE and CRPS values between MC and C, displayed against minimum great circle distance between a prediction location and its nearest sampling location. For all distances, C outperforms MC. Moreover, the differences of two prediction errors between two models increase as minimum distances between prediction locations and their nearest sampling locations increase.
On the other hand, Figure~\ref{fig:geoT_LS_data}(a) presents residual fields after removing mean structure by using simple harmonic regression depending on latitude as in \eqref{eq:mlat2}. We fitted $C^{4}$-Wendland covariance functions with the Euclidean distance (WC) and the great circle distance (WG) in addition to the covariance models considered previously with constant mean structure. For WC, the covariance function is defined by 
$$\varphi(t)=\sigma^{2}\{ 1+({t}\tau)/({cR})+{t^{2}}({\tau^{2}-1})/({3}{c^{2}R^{2}}) \}\{1-{t}/({cR})\}_{+}^{\tau},$$ where $\tau\geq 6$, $c>0$, and $t>0$. Sampling and prediction locations remained the same as in the previous example. From Table~\ref{tab:geoT_LS}, all models have comparable maximum log-likelihood values except MG. This may be due to the fact that the residual field seems smooth. Regarding prediction, C and WG have better performances than MC and WC, respectively. Note that WC has much larger sample standard deviations of estimates for support and shape parameters than WG. 

When comparing the two mean structures, prediction errors for the case of a constant mean are larger than those for the case of the mean given by \eqref{eq:mlat2}. Since the geopotential height data mainly show large scale, smooth, variation depending on latitude, the mean structure using simple harmonic regression resulted in improved prediction. Overall, the covariance functions of the class $\Psi_{d}$ performed better than those of the class $\Phi_{d+1}$ regardless of mean structures. 

\subsection{Example II: vertical directional sampling design for prediction}
\label{sec:exam_hgt2}

\begin{figure}
\begin{center}
\includegraphics[height=3.25in,width=6.5in]{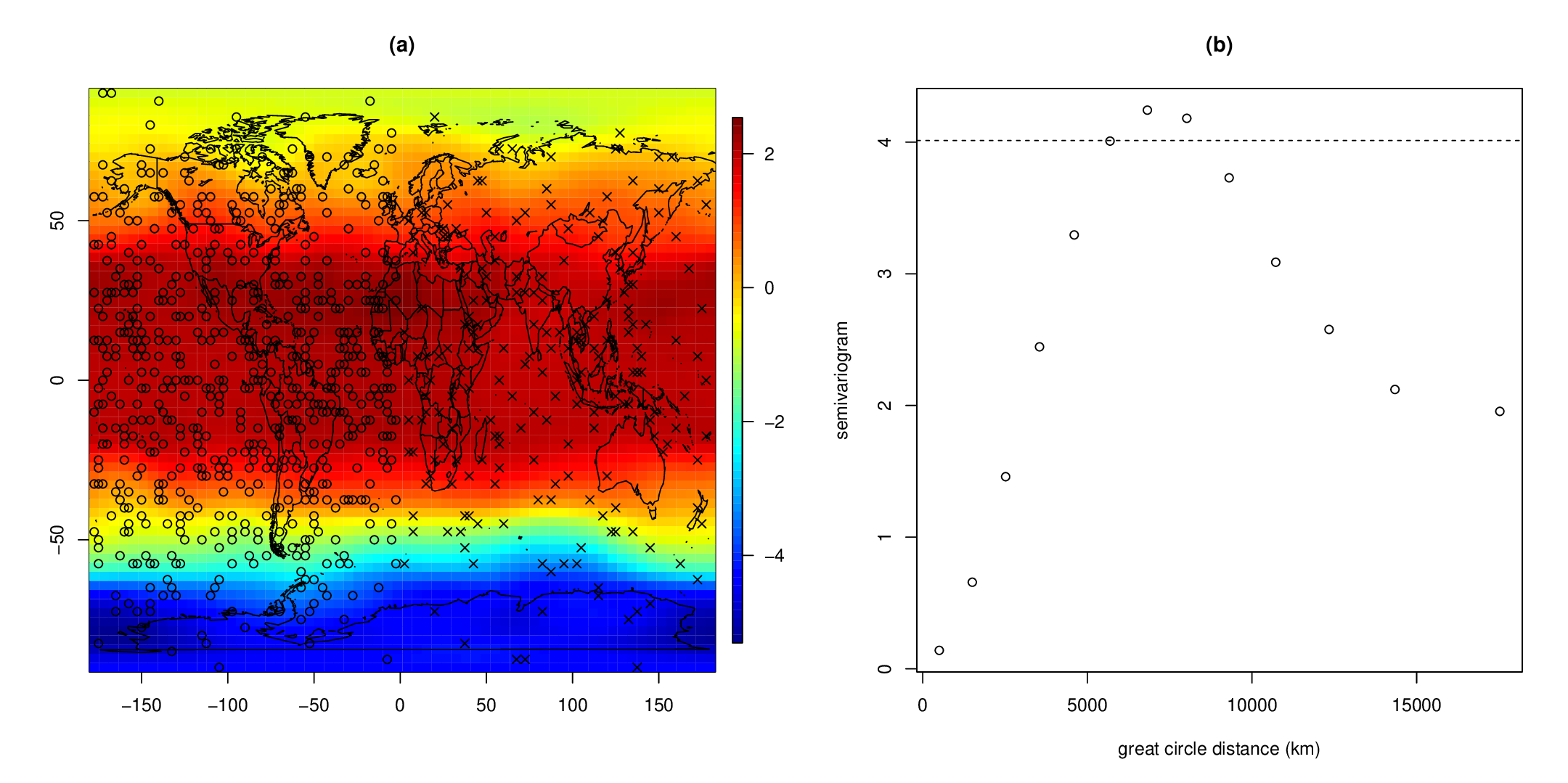}
\caption{\label{fig:geoT_half_data}(Data example II) (a) A realization of residual fields after subtracting the constant mean. For (a), sampling locations ($\circ$) and prediction locations ($\times$). (b) Empirical semi variogram values for selected locations displayed against the great circle distance. For (b), dotted line represents sample variance.}
\end{center}
\end{figure}

We entertain the same set of covariance models with the mean structures as Section~\ref{sec:exam_hgt1}. However, we changed a sampling design for prediction. We randomly selected 600 locations where longitude is less than $0^{\circ}$ for parameter estimation, and selected 200 locations where that is greater than $0^{\circ}$ for prediction as in Figure~\ref{fig:geoT_half_data}(a). The empirical semi variogram in Figure~\ref{fig:geoT_half_data}(b) shows that there is not much non-negative covariance values, unlike the previous example.

\begin{figure}[htbp]
\begin{center}
\includegraphics[height=3.25in,width=6.5in]{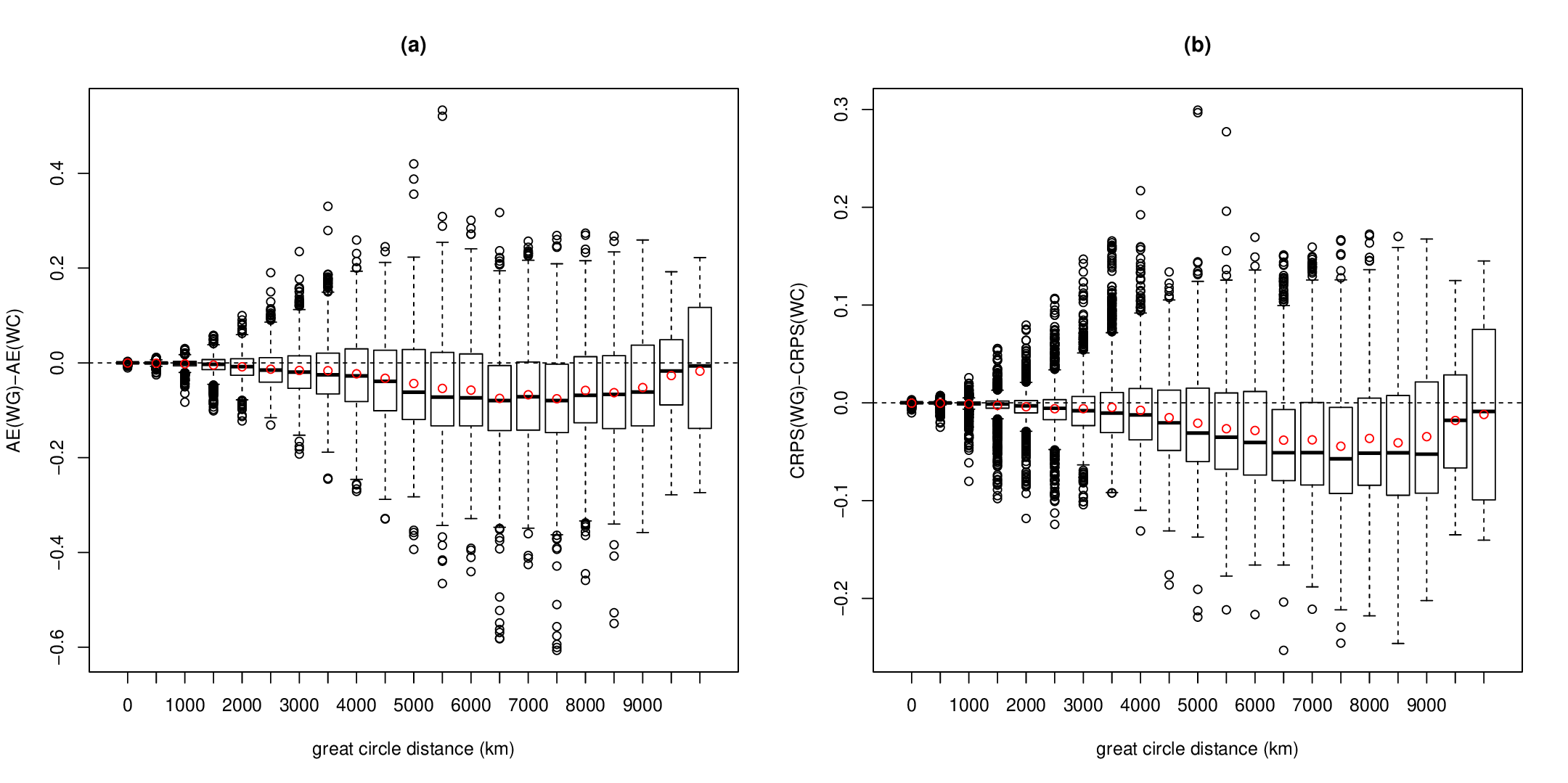}
\caption{\label{fig:geoT_half_pm3}(Data example II - the constant mean) Boxplots of differences of AE (a) and CRPS (b) values from WG and WC, displayed against minimum great circle distance between a prediction location and its nearest sampling location. Red circles represent average values in each bin.}
\end{center}
\end{figure}

With a constant mean structure, Table~\ref{tab:geoT_half} shows similar results as in Section~\ref{sec:exam_hgt1}. All models except MG are similar in terms of maximum log-likelihood values. Both C and WG have better performance in terms of prediction than MC and WC, respectively. Although results in Table~\ref{tab:geoT_half} show that models C and WG do not outperform MC and WC as significantly as in Section~\ref{sec:exam_hgt1}, there still exist some improvements in terms of prediction with the models defined with the great circle distance. From Figures~\ref{fig:geoT_half_pm3}(a) and (b), we observe that WG has smaller AE and CRPS values than WC for prediction locations who are relatively far away from their nearest sampling locations. 

When we consider the mean structure in \eqref{eq:mlat2}, there is no significant difference in terms of prediction errors between $C^{4}$-Wendland models using the great circle distance and the Euclidean distance. However, convex sum and $C^{4}$-Wendland models using the great circle distance perform better than Mat\'{e}rn covariance model using the Euclidean distance in prediction. It is expected that the vertical directional design has smaller prediction errors than the horizontal directional design in Section~\ref{sec:exam_hgt1} (see Tables~\ref{tab:geoT} and \ref{tab:geoT_half}) due to large scale variation depending strongly on latitude.

\vskip14pt
\section{Discussion}
\label{sec:dis}

We have considered several classes of isotropic covariance functions with either the great circle distance or the Euclidean distance and compared them in terms of parameter estimation and spatial prediction. We have shown that when the true spatial range is large, the prediction performance of covariance models defined on the sphere using the great circle distance (that is, $\psi(\theta)$ on $\mathcal{S}_{r}^{d}$) is better than the functions projected from the Euclidean space in spatial prediction. Moreover, when data showed significantly negative correlations at large distance lags, isotropic covariance models in the class $\Phi_{3}$ are not adequate and there is substantial difference between covariance models from the classes $\Psi_{2}$ and $\Phi_{3}$ in prediction. In geopotential height data set, we showed that the distortion of the Euclidean distance may lead to poor prediction for prediction locations that are relatively far away from sampling locations.



\vspace{1cm}
\noindent{\bf  Acknowledgment}

Mikyoung Jun's research was supported by NSF grant DMS-1208421. This publication is based in part on work supported by Award No. KUS-C1-016-04, made by King Abdullah University of Science and Technology (KAUST). The NCEP/NCAR reanalysis data used in this work was provided by the NOAA/ESRL PSD, Boulder, Colorado, USA. The authors would like to thank Peter Guttorp for discussion about this problem and Ramalingam Saravanan for suggesting to look at geopotential height data. We also thank David Bolin and H\aa vard Rue for their help to generate a oscillating Mat\'{e}rn models from the {\tt R-INLA} package.

\newpage
\vspace{1cm}
\noindent{\bf  References}

\bibliographystyle{elsarticle-harv}  
\bibliography{bib_references} 

\newpage
\begin{table}\small
\caption{\label{tab:s2}(Simulation example on $\mathcal{S}_{R}^2$) Sample means and standard deviations of parameter estimates, maximum log-likelihood values, and prediction errors for each model ($100$ cases). For model C, $\lambda\in(0,1)$ is a weighting parameter. }
\centering
\begin{tabular}{c|cccc}
\hline
Model & MC & MG & C & H\\
\hline
$\hat{\sigma}^{2}$ &6.654(1.237)& 3.260(0.448)& 26.033(7.163)& 7.496(1.711)\\
$\hat{\alpha}$ or $\hat{\lambda}$ &7918.459(1669.169)& 49423.925(7016.223)& 0.999(0.006)& 58.499(56.164)\\ 
$\hat{\nu}$ or $\hat{\beta}$ & 1.037(0.054)& 0.500(0.000)& 1.820(0.038)& $\cdot$\\
\hline
Max.loglik & 356.505(11.217)& 298.233(6.923)& 354.384(10.533)& -430.486(38.393)\\
RMSE & 1.123(0.179)& 0.975(0.085)& 0.874(0.126)& 1.961(0.369) \\
MAE & 0.903(0.177)& 0.749(0.065)& 0.717(0.118)& 1.751(0.337)\\
CRPS & 0.639(0.096)& 0.573(0.058)& 0.514(0.055)& 1.166(0.213)\\
\hline
\end{tabular} 
\end{table}

\begin{table}\small
\caption{\label{tab:geoT}(Data example I - constant mean) Sample means and standard deviations of parameter estimates, maximum log-likelihood values, and prediction errors for each model ($100$ cases). For model C, $c\in(0,\pi]$ is a support parameter and $\tau\geq 6$ is a shape parameter. }
\centering
\begin{tabular}{c|ccc}
\hline
Model & MC & MG & C\\
\hline
$\hat{\sigma}^{2}$ & 1.410(0.087)& 0.800(0.115)& 2.240(0.377)\\
$\hat{\alpha}$ or $\hat{\lambda}$ & 2128.225(114.514)& 136377.194(18394.328)& 0.811(0.108)\\ 
$\hat{\nu}$ or $\hat{c}$ & 2.605(0.072) & 0.500(0.000)& 2.931(0.084)\\
$\hat{\tau}$ & $\cdot$& $\cdot$& 7.826(0.302)\\
\hline
Max.loglik & 2167.401(20.798)& 1413.049(11.086)& 2160.817(20.348)\\
RMSE &  3.972(0.137)& 6.939(0.086)& 3.292(0.186)\\
MAE & 3.853(0.142)& 6.840(0.090)& 3.168(0.193)\\
CRPS & 3.427(0.145)& 6.767(0.089)& 2.650(0.196)\\
\hline
\end{tabular} 
\end{table}

\begin{table}\small
\caption{\label{tab:geoT_LS}(Data example I - simple harmonic regression depending on latitude) Sample means and standard deviations of parameter estimates, maximum log-likelihood values, and prediction errors for each model ($100$ cases). For models C and WG, $c\in(0,\pi]$ is a support parameter and $\tau\geq 6$ is a shape parameter. For model WG, $c>0$ is a support parameter.}
\centering
\begin{tabular}{c| ccccc}
\hline
Model & MC & MG & C & WG & WC\\
\hline
$\hat{\sigma}^{2}$ & 0.776(0.076)& 0.100(0.001)& 0.820(0.053)& 0.706(0.036)& 0.770(0.139)\\
$\hat{\alpha}$ or $\hat{\lambda}$ & 1770.346(121.073)& 10754.167(91.314)& 0.798(0.056)& $\cdot$& $\cdot$\\ 
$\hat{\nu}$ or $\hat{c}$ & 2.699(0.088)& 0.500(0.000)& 2.887(0.074)& 2.873(0.020)& 9.794(22.488)\\
$\hat{\tau}$ & $\cdot$& $\cdot$& 9.426(0.225)& 9.237(0.198)& 29.986(68.228)\\
\hline
Max.loglik & 2171.231(21.331)& 1242.183(6.178)& 2168.634(6.178)& 2168.689(20.324)& 2168.653(20.271)\\
RMSE &   2.441(0.137)& 2.351(0.057)& 2.177(0.085)& 1.988(0.074)& 2.184(0.357)\\
MAE &  2.314(0.142)& 2.217(0.065)& 2.047(0.090)& 1.856(0.079)& 2.053(0.365)\\
CRPS & 1.952(0.135)& 2.100(0.064)& 1.677(0.084)& 1.493(0.072)& 1.689(0.351)\\
\hline 
\end{tabular}  
\end{table}

\begin{table}\small
\caption{\label{tab:geoT_half}(Data example II - constant mean) Sample means and standard deviations of parameter estimates, maximum log-likelihood values, and prediction errors for each model ($100$ cases). For models C and WG, $c\in(0,\pi]$ is a support parameter and $\tau\geq 6$ is a shape parameter. For model WG, $c>0$ is a support parameter.}
\centering
\begin{tabular}{c| ccccc}
\hline
Model & MC & MG & C & WG & WC\\
\hline
$\hat{\sigma}^{2}$ & 2.062(0.278)& 0.311(0.005)& 3.621(0.315)& 3.545(0.302)& 3.249(0.208)\\
$\hat{\alpha}$ or $\hat{\lambda}$ & 1220.475(158.687)& 10527.394(87.840)& 0.934(0.084)& $\cdot$& $\cdot$\\ 
$\hat{\nu}$ or $\hat{c}$ & 3.412(0.229)& 0.500(0.000)& 2.939(0.168) & 3.026(0.061)& 3.018(0.957)\\
$\hat{\tau}$ & $\cdot$& $\cdot$& 8.539(0.648)& 8.688(0.377)& 8.783(2.826)\\
\hline
Max.loglik & 1959.170(40.787)& 921.716(12.424)& 1923.248(31.613)& 1923.297(31.593)& 1924.991(32.035)\\
RMSE &  1.167(0.079)& 1.610(0.063)& 1.107(0.118)& 1.042(0.090)& 1.068(0.097)\\
MAE & 0.929(0.075)& 1.340(0.063)& 0.855(0.105)& 0.804(0.086)& 0.832(0.090)\\
CRPS & 0.606(0.054)& 1.167(0.059)& 0.558(0.064)& 0.527(0.053)& 0.541(0.057)\\
\hline
\end{tabular} 
\end{table}

\end{document}